\def   \ni {\noindent}

\def   \ssk {\vskip  5truept}

\def   \bsk {\vskip 15truept}
 
\def   \newpage {\vfill\eject}
\def   \newline {\hfil\break}

\documentstyle[psfig]{article}
\begin{document}

\hsize 5truein
\vsize 8truein
\font\abstract=cmr8
\font\keywords=cmr8
\font\caption=cmr8
\font\references=cmr8
\font\text=cmr10
\font\affiliation=cmssi10
\font\author=cmss10
\font\mc=cmss8
\font\title=cmssbx10 scaled\magstep2
\font\alcit=cmti7 scaled\magstephalf
\font\alcin=cmr6 
\font\ita=cmti8
\font\mma=cmr8
\def\ref{\par\noindent\hangindent 15pt}
\null
%\vskip 3.0truecm
%\baselineskip = 12pt

% beginning of font "title"

\title{\ni DO NORMAL GALAXIES HOST A BLACK HOLE? THE HIGH ENERGY
PERSPECTIVE}
\footnote{To appear in the proc. of "The 3rd INTEGRAL Workshop: 
The Extreme Universe}
%(Astrophysical Letters \& Communications, 1999)
%Vol. 37

% beginning of font "author and affiliation"
\bsk \bsk
\author{\ni Y. Terashima\footnote{Present address: NASA Goddard Space Flight Center, Code 662, Greenbelt, MD 20771, USA}}
\bsk
\affiliation{ Nagoya University
}                                                
\bsk
\baselineskip = 12pt

% beginning of font "abstract and keywords"
\abstract{ABSTRACT \ni

We review ASCA results on a search for low luminosity active nuclei at
the center of nearby normal galaxies. More than a dozen low-luminosity
AGN have been discovered with 2--10 keV luminosity in the range
10$^{40-41}$~ergs~s$^{-1}$. Their X-ray properties are in some
respects similar to those of luminous Seyfert galaxies, but differ in
other respects. We also present estimated black hole masses in low
luminosity AGNs and a drastic activity decline in the nucleus of the
radio galaxy Fornax A. These results altogether suggest that relics of
the past luminous AGNs lurk in nearby normal galaxies.

}                                                    
\bsk
\baselineskip = 12pt
\keywords{\ni KEYWORDS: Galaxies; Low luminosity AGNs; LINERs; Black holes\\
}

\bsk
\baselineskip = 12pt

% beginning of font "text"

%
%def '98.9.4
% 
\newcommand{\NH}{{$N_{\rm H}$}}

\newcommand{\LX}{{$L_{\rm X}$}}
\newcommand{\LB}{{$L_{\rm B}$}}
\newcommand{\LHa}{{$L_{\rm H\alpha}$}}

\newcommand{\eps}{ergs s$^{-1}$}
\newcommand{\pcm}{cm$^{-2}$}

\text{\ni 1. Introduction
\ssk
\ni     

  The number density of quasars is peaked at a redshift of $z\sim2$
and rapidly decreases toward smaller redshifts. In the local universe,
there is no AGN emitting at huge luminosity like quasars.  These facts
infer that quasars evolve to supermassive black holes in nearby 
apparently normal galaxies (e.g. Rees 1990).

  The growing evidence for supermassive black holes in nearby galaxies
are obtained from recent optical and radio observations of gas/stellar
kinematics around the center of galaxies (e.g. Ho 1998a; Magorrian et
al. 1998; Kormendy \& Richstone 1995). If fueling to the supermassive
black hole takes place with a small mass accretion rate, they are
expected to be observed as very low luminosity AGNs compared to
quasars.

%The bulge mass - black hole mass correlation is surprisingly similar
%$M_{\rm bulge}$-$M_{\rm BH}

Recent optical spectroscopic surveys have shown that low level
activity is fairly common in nearby galaxies. In particular, a number
of LINERs (low ionization nuclear emission line regions; Heckman
1980), which are characterized by strong low ionization optical
emission lines such as [NII]$\lambda$6583, [SII]$\lambda\lambda$6716,
6731, and [OI]$\lambda6300$, are detected in nearby bright galaxies
($\sim$40\% of northern galaxies with $B_T\leq12.5$; Ho et al. 1997a, b,
c). Low luminosity Seyfert galaxies are also detected in 10\% of the
same galaxy sample. These galaxies with low level activity are good
candidates of relics of quasars ("dead" or "dormant" quasars) with a
supermassive black hole at the center.

%compact radio cores (Ho 1998; Sadler ...)

%Alternatively, galaxies showing such weak activity could be a
%distinct class of AGN with a small central engine.

  A remarkable example is NGC 4258. Makishima et al. (1994) discovered
a low luminosity AGN with a X-ray luminosity $4\times10^{40}$~{\eps}
in this galaxy. The 
mass 
\newpage
\noindent of the central black hole is measured as
$3.6\times10^7 M_{\odot}$ by H$_2$O mega maser observations (Miyoshi
et al. 1995). These observations proved that the luminosity is indeed
quite low: $\sim10^{-5}$ of the Eddington luminosity.

%  Thus detection of a compact hard X-ray source at the center is one
%of the most convincing evidence for AGN. 

  We present results of X-ray observations of nearby galaxies and
discuss the origin of low level activity and show low luminosity AGNs
(hereafter LLAGNs) are numerous in the local universe. We also
summarize X-ray properties of LLAGNs and attempt to set a limit on the
central black hole mass. An example of a drastic decline of activity
in the radio galaxy Fornax A is also presented.

\bsk
\ni 2. A search for AGN in nearby galaxies
%\ssk
%\ni

\bsk
\ni
2.1 Low level activities in nearby galaxies
\ssk
\ni

Optical spectroscopic surveys revealed that LINERs are very numerous.
The origin of LINERs are still under debate and several mechanisms are
proposed to explain LINER optical emission lines, such as
photoionization by LLAGNs, photoionization by very hot stars, shocks,
and so on (see Filippenko 1996 for a review). Recent observational
progress in various wavelengths provide pieces of evidence for LLAGNs
in at least some fraction of LINERs (Ho 1998b and references
therein). X-ray observations are one of the most important tools to
search for LLAGNs and we summarize the results of {\it ASCA}
observations in the following subsections.

\bsk
\ni
2.2 ASCA observations of LINERs and low luminosity Seyfert galaxies
\ssk
\ni

We analyzed X-ray data of 17 LINERs obtained with {\it ASCA} and also
10 low luminosity Seyfert galaxies (hereafter LLSeyferts) with
H$\alpha$ luminosities less than $\sim 10^{40}$ {\eps}. Optical
classifications are adopted from Ho et al. (1997a) except for NGC 1097
(Storchi-Bergman et al. 1993) and IC 1459 (Philipps et al. 1986). The
observed galaxies and optical classifications are summarized in Table
1.

\begin{table*}
\begin{center}
\caption{Table 1. Observed galaxies}
\begin{tabular}{ll}
\hline\hline
classification  & name\\
\hline
LINER 1         & NGC 315, NGC 1052, NGC 1097, NGC 3998, NGC 4203, NGC 4438\\
                & NGC 4450, NGC 4579, NGC 4594, NGC 5005, IC 1459\\
LINER 2         & NGC 404, NGC 4111, NGC 4569, NGC 4736, NGC 5195, NGC 7217\\
Seyfert 1       & NGC 2639, NGC 3031, NGC 4258, NGC 4565, NGC 4639, NGC 5033\\
Seyfert 2       & NGC 2273, NGC 3079, NGC 3147, NGC 4941\\
\hline
\end{tabular}
%\begin{list}{Table Notes.}
%\item $^a$ calculated using H$_0$=75 km s$^{-1}$ Mpc$^{-1}$
%\end{list}
\end{center}
\end{table*}

Since the X-ray continuum shape of LLAGNs are quite similar to
luminous AGNs as discussed in the next section, we searched for a
point-like X-ray source with an AGN-like continuum at the center of
LINER-type galaxies. We detected X-ray sources from all LINERs but for
NGC 404. X-ray spectra are well represented by a two component model:
a power-law with a photon index of $\Gamma \sim 1.7-2$ suffered from
photo electric absorption ({\NH}$\sim10^{20}-10^{23}$~{\pcm}) plus a
soft thermal component with a temperature of $kT\sim0.5-1$ keV. The
hard component of most of the galaxies can be also represented by a
thermal bremsstrahlung model with a temperature of several keV. X-ray
luminosities of the hard component range from $4\times10^{39}$~{\eps}
to $6\times10^{41}$~{\eps}. Examples of {\it ASCA} spectra of LINERs
are shown in Figure 1.

%Firstly we show two examples of low luminosity Seyfert 1.5 NGC 5033
%and Seyfert 2 NGC 4941 in Figure 1., since it is not obvious whether
%X-ray properties of LLAGNs are same as luminous AGNs or not. The X-ray
%spectrum of NGC 5033 is well represented by a power-law with a photon
%index of 1.7 suffered from small absorption ($\NH\sim9\times10^{20}$
%{\pcm})....

\begin{figure}[t]
%\centerline{
%\hspace{1.0cm}
%\psfig{file=n5033.ps,width=6.5cm,height=5cm,angle=-90}
%\hspace{-1.5cm}
%\psfig{file=n4941.ps,width=6.5cm,height=5cm,angle=-90}
%}
%\end{figure}
\vspace{-1.5cm}
%\begin{figure}
\centerline{
\hspace{1.0cm}
\psfig{file=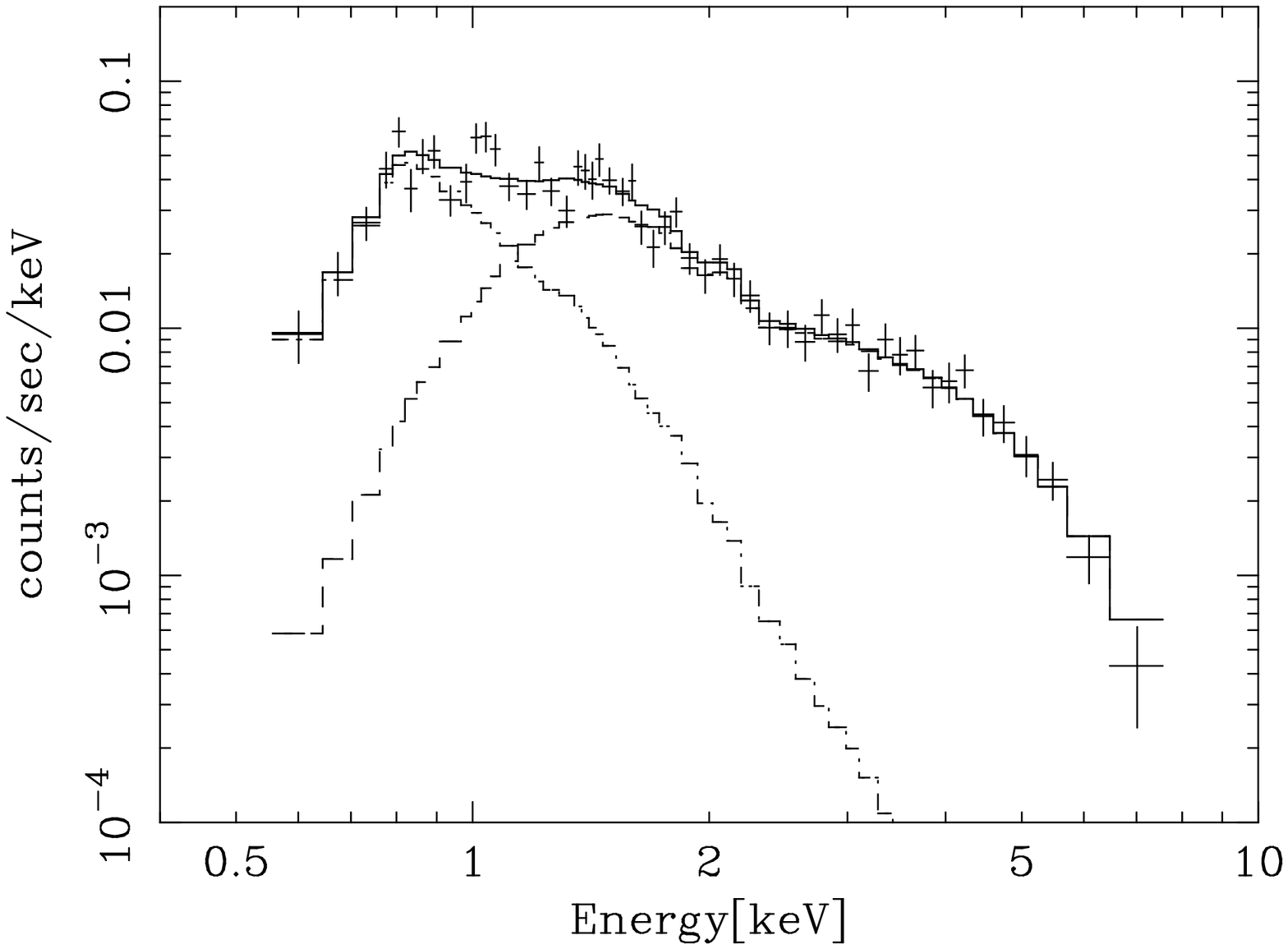,width=7.5cm,height=6cm,angle=-90}
\hspace{-1.5cm}
\psfig{file=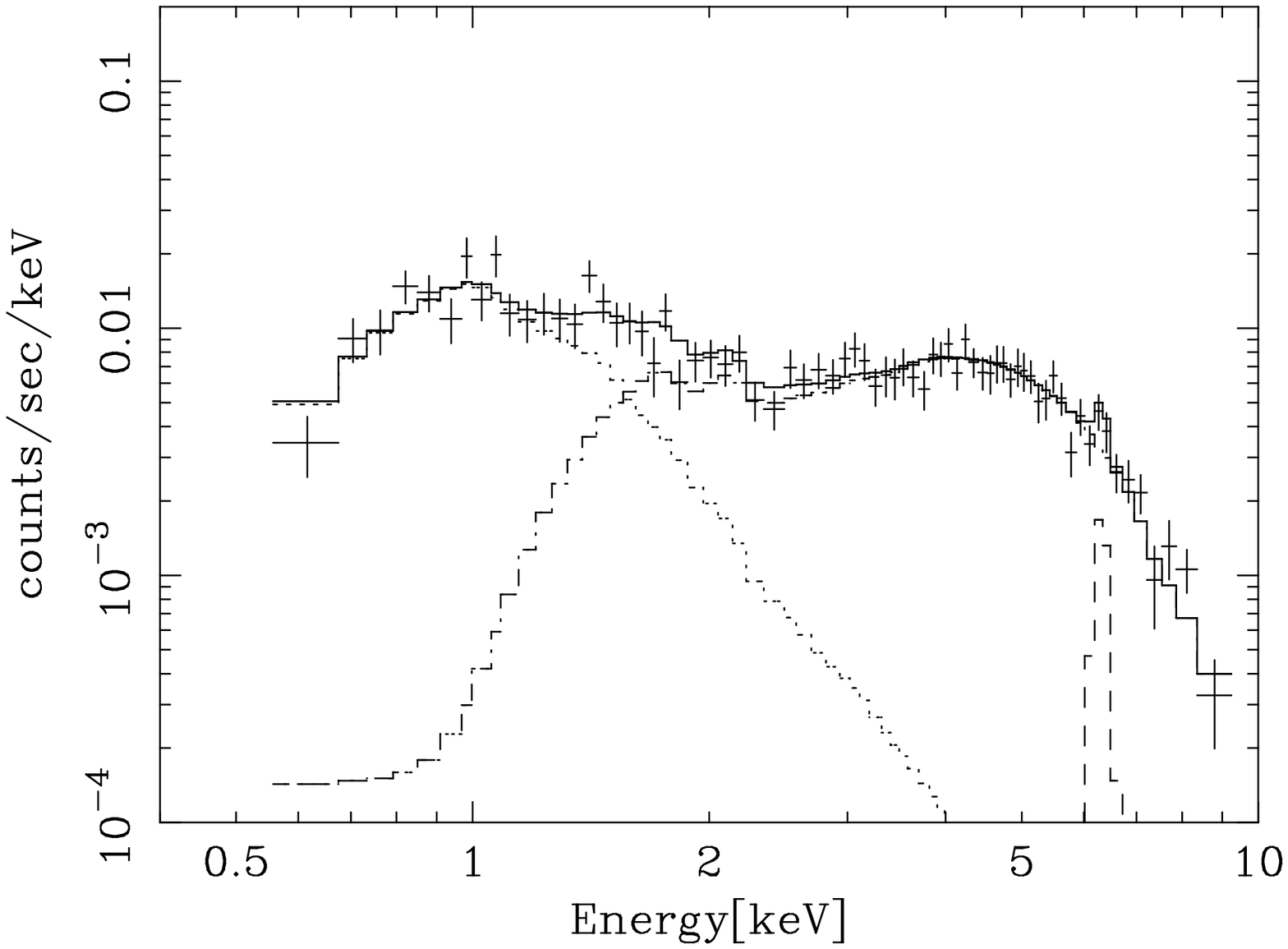,width=7.5cm,height=6cm,angle=-90}
}
\vspace{-0.5cm}
%\caption{FIGURE 1. ASCA SIS spectra of (a) low luminosity Seyfert 1.5
% NGC 5033, (b) low luminosity Seyfert 2 NGC 4941, 
%(c) LINER 1.9 NGC 4594, and (d) LINER 1.9 with polarized broad H$\alpha$ NGC 1052. }
\caption{FIGURE 1. Examples of ASCA SIS spectra ({\it left}) LINER 1.9 NGC 4594, and ({\it right}) LINER 1.9 with polarized broad H$\alpha$ NGC 1052. }
\end{figure}

\bsk
\ni
2.3 Ionizing source of LINERs
\ssk
\ni

Hard X-ray emission from galaxies can be produced by various origins
such as AGN, X-ray binaries, and starburst activity. If LLAGNs are the
dominant energy source of X-ray emission and optical LINER emission
lines, optical emission line luminosities are expected to be
proportional to X-ray luminosities as is observed in luminous Seyfert
1 galaxies and quasars (e.g. Ward et al. 1988). Figure 2 shows the
correlation between X-ray luminosities ({\LX}) and H$\alpha$
luminosities ({\LHa}: total of broad and narrow H$\alpha$) for LINERs
with broad H$\alpha$ (hereafter LINER 1s) and Seyfert 1s. Data for
luminous Seyfert 1s and quasars are taken from Ward et al. (1988). The
correlation extends to lower luminosities and strongly supports 
the hypothesis that the
primary ionizing source of LINER 1s is LLAGNs. Note that the
{\LX}/{\LHa} for starburst galaxies are about 2 orders of magnitudes
smaller than Seyfert galaxies (P\'erez-Olea \& Colina 1996).

% and shock excitation by starburst driven winds is not a dominant
%excitation mechanism in the present sample.

% LINER 2s

Thus the hard X-ray emission from LINER 1s is most likely to be
dominated by LLAGN. Then we compare luminosity ratios {\LX}/{\LHa} of
LINERs without broad H$\alpha$ (hereafter LINER 2s) with those of
LINER 1s and LLSeyferts. Histograms of {\LX}/{\LHa} values for LINER
1s + LLSeyferts and LINER 2s are shown in Figure 2. The {\LX}/{\LHa}
values for LINER 2s are systematically lower than LINER 1s and
LLSeyferts except for NGC 4736, which has a {\LX}/{\LHa} ratio similar
to LINER 1s + LLSeyferts.

If we assume a spectral energy distribution from UV to X-ray in the
form of a power-law with an index of --1 ($f_{\nu} \propto \nu^{-1}$),
Case B recombination (Osterbrock 1989) and a covering fraction of 1,
the objects with low {\LX}/{\LHa} are likely to be too X-ray weak to
ionize optical emission lines. If an AGN is present in these objects,
such AGN should be obscured even at energies above 2
keV. Alternatively there might be other ionization sources. Actually
UV spectra of NGC 4569 and NGC 404 obtained with $HST$ Faint Object
Spectrograph (FOS) show the presence of hot stars (Maoz et al. 1998).

%{\LX}/{\LHa} should be greater than XXX to drive H$\alpha$ luminosity
%by photoionization under 

%Since X-ray luminosities of LINER 2s are small
%({\LX}$<\sim 2\times10^{40}$), X-ray binaries in the host galaxy
%might also contribute to observed X-ray fluxes. Then {\LX}/{\LHa}
%value for nuclear X-ray emission becomes smaller. Therefore 

\begin{figure}[t]
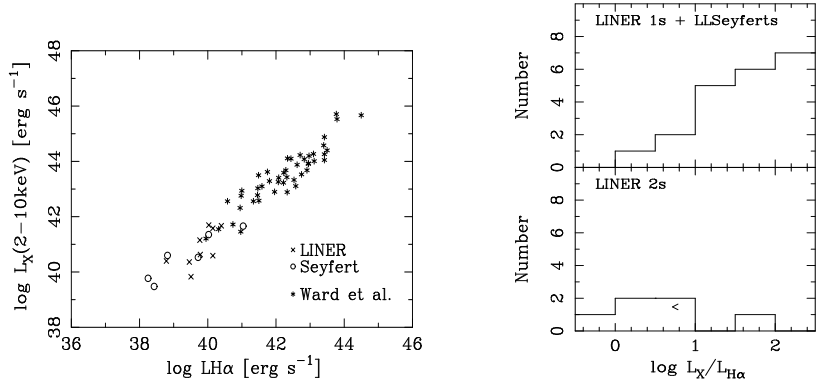

\centerline{
\hspace{0.3cm}
\psfig{file=Ha.ps,width=5.5cm,height=4.5cm,angle=-90}
\hspace{1.0cm}
\psfig{file=hist.ps,width=4.0cm,height=5cm,angle=-90}
}
\vspace{0.5cm}
\caption{FIGURE 2. ({\it left}) Correlation between X-ray and H$\alpha$
luminosity for LINER 1s and Seyfert 1s.
({\it right}) X-ray to H$\alpha$ luminosity ratio for LINER 1s + 
LLSeyfert ({\it upper}) and LINER 2s ({\it lower}).
}
\end{figure}

%\bsk
%\ni
%2.4 AGN fraction of nearby galaxies
%\ssk
%\ni

According to the optical spectroscopic survey by Ho et al. (1997a),
Seyferts, LINER 1s and LINER 2s are detected in 11\%, 5\%, 28\% of
northern bright galaxies. If we assume the most extreme case that all
Seyferts and LINER 1s are AGNs and all LINER 2s are not AGNs, we
estimate the fraction of AGNs in bright galaxies is $\sim$16 \%. Since
a certain fraction of LINER 2s is probably genuine AGNs, this
percentage should, however, be regarded as a lower limit. Therefore
the number density of AGNs in the local universe is considered to be
larger than previously thought.

%There are LINER 2s with AGN

%NGC 4486 (M87)

%NGC 4261

%Therefore a certain fraction of LINER 2s are actually AGN. A
%conservative lower limit of the AGN fraction in nearby galaxies is
%calculated by assuming (1) all Seyferts and LINERs with broad
%H$\alpha$ are AGNs, (2) almost all Seyfert 2s are AGNs, and (3) all
%LINER 2s are not AGNs. 

%The lower limit XXX \% 

%\newpage

\bsk
\ni 3. X-ray properties of low luminosity AGNs and black hole mass}
%\ssk
%\ni     

\bsk
\ni
3.1 X-ray spectra
\ssk
\ni

We summarize X-ray properties of LLAGNs using galaxies from which
X-ray emission is dominated by AGNs. X-ray images of these galaxies
are consistent with point like in the hard X-ray band and {\LX}/{\LHa}
ratios are also similar to luminous Seyferts. We use data of (1) NGC
315, 1097, 3031, 3079, 3147, 3998, 4203, 4450, 4579, 4594, 4639, 4736,
5033, IC 1459 and (2)NGC 1052, 2273, 2639, 4258, 4941. The absorption
column densities for galaxies in the group (1) and group (2) are
{\NH}$<10^{23}$ {\pcm} and {\NH}$>10^{23}$ {\pcm}, respectively. We
classify these subclasses as Type 1 and Type 2 LLAGNs, respectively.

  The luminosity dependence of photon indices for Type 1 AGNs is shown
in Fig 3. The data for luminous Seyfert 1 galaxies are taken from
Nandra et al. (1997b). The histogram of absorption column densities is
also shown in Fig 3. These figures indicate that X-ray spectra of Type
1 LLAGNs have quite similar photon indices to luminous AGNs. The
photon indices of Type 2 LLAGNs also show no luminosity dependence,
although errors for indices are large for both luminous and low
luminosity Type 2 AGNs. The absorption column densities range from
$10^{20}$ {\pcm} to $10^{24}$ {\pcm} and there exists LLAGNs both with
heavy absorption and small absorption as in Seyfert galaxies. Note
that the present sample is not complete and that the distribution of
absorption column densities does not reflect the optical depths and
geometry of the obscuring matter around LLAGNs.

  Thus X-ray continuum shape of LLAGNs are quite similar to luminous
AGNs. On the other hand, iron K emission line properties and
variability in Type 1 LLAGNs are different from luminous AGNs.  No
significant iron K emission is detected from our Type 1 LLAGN sample
except for NGC 3031, NGC 4579, and NGC 5033, in contrast to 
Seyfert 1s which generally show iron K emission at 6.4 keV. The center
energy of the iron K lines in NGC 3031 and NGC 4579 are 6.7 keV
(Ishisaki et al. 1996; Terashima et al. 1998a), which is significantly
higher than Seyfert 1s, while that of NGC 5033 is 6.4 keV (Terashima
et al. 1998b). These iron line properties are possibly due to
difference of physical states of accretion disks between LLAGNs and
luminous AGNs.

%continuum: photon indices / absorption column densities

%iron K emission : Sy 2

%iron K emission : Sy 1

%315, 1097, 3998, 4203, 4450, 4594, IC 1459

\begin{figure}[t]
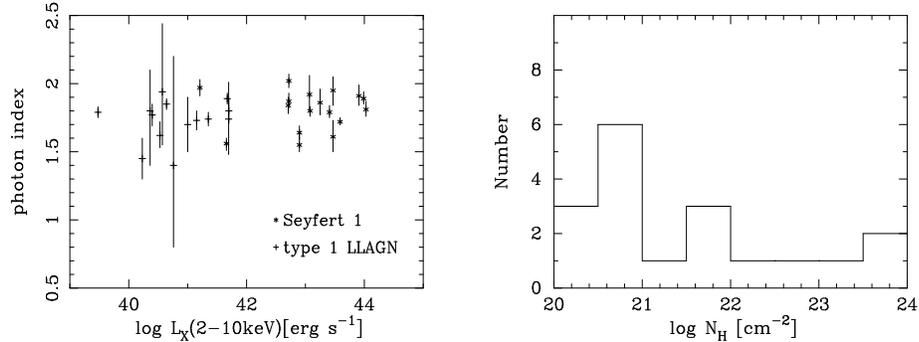

\centerline{
\hspace{0.3cm}
\psfig{file=index.ps,width=5.5cm,height=4.5cm,angle=-90}
\hspace{0.7cm}
\psfig{file=abs.ps,width=5.5cm,height=4.5cm,angle=-90}
}
\caption{FIGURE 3. ({\it left}) Luminosity dependence of photon indices
 of type 1 AGNs. ({\it right}) Distribution of absorption column densities of LLAGNs
}
\end{figure}

\bsk
\ni
3.2 X-ray variability and black hole mass
\ssk
\ni

Rapid time variability on time scales less than one day is generally
observed in luminous Seyfert 1 galaxies. It is well known that lower
luminosity Seyfert 1 galaxies tend to show larger amplitude and
shorter time scale variability (Lawrence \& Papadakis 1993; Nandra et
al. 1997a).

We searched for short-term variability on time scales less than one
day for the galaxies in the present sample. Only two low luminosity
Seyfert 1 galaxies (NGC 3031 and NGC 5033) indicate variability on
time scale of $\sim10^4$ sec (Ishisaki et al. 1996; Terashima et
al. 1998b). Thus small amplitude or no detectable variability is quite
common in LLAGNs, and luminosity dependence of variability properties
observed in luminous Seyfert 1s are no longer seen in LLAGN with
luminosities below $\sim5\times10^{41}$ {\eps}.

%Several objects are also observed by {\it Ginga} or {\it
%ROSAT}. Comparison with these data indicates long term variability on
%time scales larger than a year.

% in NGC 3031, 3998, 4258, 4579, 5033.

One possibility of such difference between luminous and less luminous
AGNs is that the structure of an accretion disk is different between
these two classes (Ptak et al. 1998). Alternatively, if the
variability time scale reflects the size of the X-ray emitting region,
the variability on longer time scales would infer larger system size
and larger black hole mass.

%--- [$L$ dependence / hot spot / mdot .....]

%supermassive black hole?

We attempt to measure black hole masses in LLAGNs from X-ray
variability using the method by Hayashida et al. (1998).  Hayashida et
al. (1998) defined a new variability time scale: the frequency at
which (normalized power spectral density) $\times$ (frequency) crosses
a certain level, where a normalized power spectral density is defined
as the power spectral density divided by the averaged source intensity
squared. If we assume that the black hole masses are linearly
proportional to the variability time scale and that the mass of Cyg
X-1 is 10$M_{\odot}$, which is used as a reference point, we can
estimate the black hole mass in AGNs.

  We estimate the black hole masses from {\it ASCA} light curves of
bright LLAGNs (NGC 1097, 3031, 3998, 4579, 5033, 4258) under an
additional assumption: the power spectral density for LLAGNs are same
as luminous AGNs, i.e. the power-spectral slope is
$\alpha\approx2$. We obtained lower limit for the black hole masses
similar to or larger than Seyfert 1 galaxies analyzed by the same
method (MCG--6--30--15, NGC 4151, NGC 5548, etc. ; Hayashida et
al. 1998) (Awaki et al. 1998). The lower limit of the black hole
masses obtained here are consistent with those obtained by other
method (NGC 3031, NGC 4258, NGC 4579). Since X-ray luminosities of
LLAGNs are 1--3 orders of magnitude smaller than luminous Seyfert 1
galaxies, the Eddington ratios ($L_{\rm bolometric}/L_{\rm Eddington}$)
of LLAGNs are estimated to be at least 1--3 orders of magnitude
smaller than luminous AGNs. These results support the idea that LLAGNs
have a black hole with a huge mass but radiating at a very low
Eddington ratio.

%Note that it is not obvious whether the accretion under such extremely
%low Eddington ratio is same as usual Seyferts or not.

\bsk
\ni 4. Declined Activity in the Nucleus of Fornax A
\ssk
\ni

The decline of the space density of quasars indicate that AGN activity
lasts shorter than the cosmological time scale. A remarkable
example of a decline of activity is the radio galaxy NGC 1316 (Fornax
A) (Iyomoto et al. 1998).

NGC 1316 is a radio galaxy with prominent radio lobes (e.g. Ekers et
al. 1983) which indicate Fornax A was active in the past. However the
nucleus of NGC 1316 is very inactive in radio and X-ray at present, and
this suggests that the nucleus is currently inactive and that the
lobes are a relic of the past activity. We summarize observational
facts and estimate various time scales of the lobes and decline of
activity according to Iyomoto et al. (1998).

\newpage

\bsk
\ni
4.1 The Nucleus
\ssk
\ni

The nucleus of NGC 1316 is quite faint in the radio band. The ratio of
the core flux to the total (lobe + core) flux is about two orders of
magnitude smaller than the typical value for radio galaxies.  A {\it
ROSAT} HRI failed to detected an X-ray nucleus (Kim, Fabbiano, \&
Mackie 1998). An {\it ASCA} spectrum is consistent with normal
elliptical galaxies, i.e. a hot plasma of $kT\sim1$~keV and hard
component presumably due to discrete X-ray sources in the host galaxy,
and its X-ray luminosity in the 2--10 keV is
$1.3\times10^{40}$~{\eps}.  An upper limit of the AGN contribution to
the hard X-ray emission is calculated to be $2\times10^{40}$~{\eps} by
assuming various absorbing column densities ($<10^{24}$~{\pcm}) for
the AGN component. If AGN is completely obscured in the energy band
below 10 keV, only the scattered component might be expected. In this
case, the intrinsic luminosity is constrained to be less than
$1.4\times10^{41}$~{\eps}, if an upper limit of iron K emission and
assumed scattering fraction $\sim$ 1\% are taken into account.

\bsk
\ni
4.2 The Lobes
\ssk
\ni

The inverse Compton X-rays, which are produced when the relativistic
electrons scatter off the cosmic microwave background photons, are
observed from radio lobes of NGC 1316. Comparing the radio and X-ray
brightnesses, the magnetic field strength in the lobes has been
determined to be 3 $\mu$G (Feigelson et al. 1995; Kaneda et al. 1995).
The lobe radio spectrum extends at least to 5 GHz without a
cutoff. Then the electron emitting 5 GHz photons lose half their
energy on a characteristic time scale of $\tau\sim0.08$ Gyr, which can
be regarded as the synchrotron life time of the lobes. Thus the nucleus
must have been active until at least $\tau$ years ago, or even until a
more recent epoch.

The X-ray luminosity that the NGC 1316 nucleus used to have in the
past can be estimated to be $\sim4\times10^{41}$~{\eps} from the
kinetic energy of the radio lobes, a correlation between kinetic
luminosities and optical narrow line luminosities, and a correlation
between [OIII] and X-ray luminosities. Then the present activity is at
least an order of magnitude lower than the estimated past activity and
the nucleus of NGC 1316 has become dormant during the last 0.1 Gyr.
This suggests the possible abundance of "dormant" quasars in nearby
galaxies.

%} % end of text

\bsk
\ni 5. Summary
\ssk
\ni     

  We have detected many LLAGNs in nearby galaxies and show that a
number of AGNs in the local universe is significantly larger than that
previously thought. Our estimation of the black holes in LLAGNs
indicates that these LLAGNs host a supermassive black hole and radiate
at a very low Eddington ratio. A drastic decline of activity is
observed in the nucleus of NGC 1316. These facts suggest that relics
of the past luminous AGNs lurk in the nearby normal galaxies.

\newpage

\bsk
\baselineskip = 12pt
{\abstract \ni ACKNOWLEDGMENTS
The author acknowledges his collaborators
K. Makishima, N. Iyomoto, H. Awaki, 
H. Kunieda, K. Misaki, Y. Ishisaki,
L.C. Ho, A.F. Ptak, 
P.J. Serlemitsos, and R.F. Mushotzky.
%ASCA team

}

\bsk
\baselineskip = 12pt

% beginning of font "references"

{\references \ni REFERENCES
\ssk

\ref Awaki, H. et al. 1998, in preparation

\ref Ekers, R.D. et al. 1983, A\&A, 127, 361

\ref Feigelson, E.D., Laurent-Muehleisen, S.A., Kollgaard, R.I., \&
Fomalont, E.B. 1995, ApJ, 449, L149

\ref Filippenko, A.V. 1996, in The Physics of LINERs in View of Recent
Observations, ed. M. Eracleous et al. (San Fransisco: ASP)

\ref Hayashida, K., Miyamoto, S., Kitamoto, S., Negoro, H., \& Inoue,
H. 1998, ApJ, 500, 642

%\ref Hawkins, M.R.S. \& Veron, P. 1995, MNRAS, 275, 1102

%\ref Hawkins, M.R.S. \& Veron, P. 1996, MNRAS, 281, 348

\ref Heckman, T.M. 1980, A\&A, 87, 152, 1980		% LINER

\ref Ho, L.C., Filippenko, A.V., \& Sargent, W.L.W. 1997a, ApJS, 112, 315

\ref Ho, L.C., Filippenko, A.V., Sargent, W.L.W., \& Peng, C.Y. 1997b, ApJS, 11
2, 391

\ref Ho, L.C., Filippenko, A.V., \& Sargent, W.L.W. 1997c, ApJ, 487, 568

\ref Ho, L.C. 1998a, in Observational Evidence for Black Holes in
the Universe, ed. S.K. Chakrabarti, (Dordrecht: Kluwer), in press

\ref Ho, L.C. 1998b, in The 32nd COSPAR Meeting, The AGN-Galaxy
Connection (Advances in Space Research), in press

\ref Ishisaki, Y. et al. 1996, PASJ, 48, 237		% M81

\ref Iyomoto, N., Makishima, K., Tashiro, M., Inoue, S., Kaneda, H.,
Matsumoto, Y., \& Mizuno, T. 1998, ApJ, 503, L31 % Fornax A

\ref Kaneda, H. et al. 1995, ApJ, 453, L13

%\ref Kennefick, J.D., Djorgovski, S.G., \& Meylan, G. 1996, AJ, 111, 1816

\ref Kim, D.-W., Fabbiano, G., \& Mackie, G. 1998, ApJ, 497, 699 % Fornax A

\ref Kormendy, J. \& Richstone, D. 1995, ARA\&A, 33, 581

\ref Lawrence, A. \& Papadakis, I. 1993, ApJ, 414, L85	% variability

\ref Magorrian, J. et al. 1998, AJ, 115, 2285		% BH

\ref Makishima, K. et al. 1994, PASJ, 46, L77        	%NGC4258

\ref Maoz, D. et al. 1998, AJ, 116, 55			% FOS

\ref Miyoshi, M., Moran, J., Herrnstein, J., Greenhill, L., Nakai, N.,
Diamond, P., \& Inoue, M. 1995, Nature, 373, 127	% n4258

\ref Nandra, K., George, I.M., Mushotzky, R.F., Turner, T.J., \& Yaqoob, T. 1997a, ApJ, 476, 70 					% Sy1 I

\ref Nandra, K., George, I.M., Mushotzky, R.F., Turner, T.J., \& Yaqoob, T. 1997a, ApJ, 477, 602 					% Sy1 II

\ref P\'erez-Olea, D \& Colina, L 1996, ApJ, 468, 191

\ref Phillips, M.M., Jenkins, C.R., Dopita, M.A., Sadler, E.M., \&
Binette, L. 1986, AJ, 91, 1062				% IC 1459 

\ref Ptak, A., Yaqoob, T., Mushotzky, R., Serlemitsos, P., Griffiths,
R. 1998, ApJL, 501, L37 				% var

\ref Rees, M.J. 1990, Science, 247, 817

\ref Storchi-Bergman, T., Baldwin, J., \& Wilson, A. 1993, ApJ, 410, L11 %n1097

%\ref Terashima, Y. et al. 1998a, ApJ, 496, 210		% M51

\ref Terashima, Y., Kunieda, H., Misaki, K., Mushotzky, R.F., Ptak,
A.F., \& Reichert, G.A. 1998a, ApJ, 503, 212 % N 4579

\ref Terashima, Y., Kunieda, H., \& Misaki, K. 1998b, PASJ, submitted % N 5033

\ref Ward, M. et al. 1988, ApJ, 324, 767

}

\end{document}